%% file: Radv2.tex
\pdfoutput=1

\documentclass[aps,prl,twocolumn,superscriptaddress,groupedaddress]{revtex4} 

\usepackage{graphicx}  
\usepackage{dcolumn}   
\usepackage{bm}        
\usepackage{amssymb}   
\usepackage{amsmath}
\usepackage{slashed}
\usepackage{ytableau}
\usepackage{hyperref}
\hypersetup{colorlinks=true,linkcolor=magenta,anchorcolor=green,citecolor=blue,filecolor=black,menucolor=black,urlcolor=blue}
\usepackage[force]{feynmp-auto} 

\usepackage{cleveref}
\newcommand{\marrow}[5]{%
    \fmfcmd{style_def marrow#1
    expr p = drawarrow subpath (1/4, 3/4) of p shifted 6 #2 withpen pencircle scaled 0.4;
    label.#3(btex #4 etex, point 0.5 of p shifted 6 #2);
    enddef;}
    \fmf{marrow#1,tension=0}{#5}}
\begin{document}
\widetext

\title{From Scattering Amplitudes to Classical Physics: Universality,
Double Copy and Soft Theorems}
\input author_list.tex      
\begin{abstract}
We introduce a covariant Multipole Expansion for the scattering
of a massive particle emitting photons or gravitons in $D$ dimensions. We find that
these amplitudes exhibit very powerful features such as universality,
soft exponentiation, orbit and spin multipoles, etc. Using ${\rm{SO}}(D)$ representation theory we show
that the photon and graviton amplitudes are related via a new double copy
procedure for massive spinning states. All these features are then promoted to properties of the observables
arising in the classical version of such theories. Focusing on radiation,
we provide two main applications: 1) An exponential Soft Theorem
relating conservative effects and gravitational radiation to all orders in $\omega$; whose leading order directly leads to the $D=4$ Memory Effect. 2) A \textit{classical} double copy to evaluate gravitational radiation from QED Bremsstrahlung, matching previous classical computations and extending them to spin-quadrupole order.
\end{abstract}
\maketitle
With the advent of QFT it was observed that 
dynamics of massive objects subject to long-range forces could be described from the classical limit of Scattering Amplitudes \cite{10.1143/PTP.46.1587,PhysRevD.7.2317,Boulware1975,Barker1979,Gupta:1980zu,Donoghue:1994dn,Holstein:2004dn,Holstein:2008sw}. 
This picture has seen renewed interest with the aim of providing more accurate templates for Gravitational Wave (GW) events, leading to remarkable Post-Minkowskian (PM) results \cite{Damour:2016gwp,Guevara:2017csg,Bjerrum-Bohr:2018xdl,Damour:2017xzj,Cheung:2018wkq,Vines:2018gqi,Guevara:2018wpp,Chung:2018kqs,Bern:2019nnu,Antonelli:2019ytb}. 
Two key ingredients in this endeavour are the following amplitudes,
\begin{equation}
\begin{fmffile}{M4M5}
   M_4= \quad\parbox{40pt}{
   \begin{fmfgraph*}(32,32)
      \fmfleftn{i}{2}
       \fmfrightn{o}{2}
\fmf{plain,tension=5,width=0.7,foreground=(0.035,,0.168,,0.623)}{i1,v1}
\fmf{plain,tension=5,width=0.7,foreground=(0.035,,0.168,,0.623)}{v1,i2}
\fmf{plain,tension=5,width=0.7,arrow.size=10,foreground=(1,,0.1,,0.1)}{o1,v1}
\fmf{plain,tension=5,thin,width=0.7,foreground=(1,,0.1,,0.1)}{v1,o2}
\marrow{eb}{left}{lft}{$a$}{i1,v1}
\marrow{ea}{left}{top}{}{v1,i2}
\marrow{ec}{right}{top}{}{v1,o2}
\marrow{ed}{right}{rt}{$b$}{o1,v1}
\fmfv{decor.shape=circle,decor.filled=12,decor.size=.3w}{v1}
\end{fmfgraph*}}\,\,\,,\,\,\,\,\,\,\,\,\,\,\,\,\,\,\,
M_5=\quad\parbox{50pt}{
    \begin{fmfgraph*}(32,32)
\fmfstraight
      \fmfleftn{i}{2}
       \fmfrightn{o}{2}
       \fmftop{t}
\fmf{plain,tension=5,width=0.7,foreground=(0.035,,0.168,,0.623)}{i1,v1}
\fmf{plain,tension=3,width=0.7,foreground=(0.035,,0.168,,0.623)}{v1,i2}
 \fmf{wiggly,tension=3,width=0.7}{v1,t}
\fmf{plain,tension=5,width=0.7,arrow.size=10,foreground=(1,,0.1,,0.1)}{o1,v1}
\fmf{plain,tension=3,thin,width=0.7,foreground=(1,,0.1,,0.1)}{o2,v1}
\marrow{eb}{left}{lft}{$a$}{i1,v1}
\marrow{ea}{up}{top}{}{v1,i2}
\marrow{ec}{right}{top}{}{v1,o2}
\marrow{ed}{right}{rt}{$b$}{o1,v1}
\marrow{ea}{left}{top}{}{v1,t}
\fmfv{decor.shape=circle,decor.filled=12,decor.size=.3w}{v1}
\end{fmfgraph*}},
\end{fmffile}
\end{equation}
which are associated to conservative and non-conservative effects  \cite{Goldberger:2016iau,Kosower:2018adc}. The bodies $a$ and $b$ carrying internal structure are here understood as point particles with spin, which is especially relevant in the quest for better templates \cite{TheLIGOScientific:2016wfe}. While $M_4$ has been studied to high PM orders, $M_{5}$ is much more complicated. It has only recently been introduced in this context by O'Connell et al. in the spinless case \cite{Kosower:2018adc,Luna:2017dtq}.
Even though these objects control fundamental effects such as the Coulombian/Newtonian potentials,
both $M_{4}$ and $M_{5}$ strongly depend on the matter content even
if no contact interactions are allowed. We will argue that the reason for their classical
piece, $\langle M_{n}\rangle:{=}\lim_{\hbar\rightarrow0}M_{n}$, to
be universal is that it is precisely identified with their decomposition into  fundamental \textit{amplitudes}. The main example we provide is that, at LO in the coupling,
\small
    \begin{equation}
 \begin{fmffile}{cutm4}\langle M_4\rangle=
 \parbox{21pt}
 {
  \parbox{6pt}{\begin{fmfgraph*}(23,30)
    \fmfstraight
    \fmfleft{i1,i2}
    \fmfright{o1}
    \fmf{plain,width=0.7,label=$p_1$,foreground=(0.035,,0.168,,0.623)}{i1,v}
    \fmf{plain,width=0.7,foreground=(0.035,,0.168,,0.623)}{v,i2}
    \fmfv{decor.shape=circle,decor.filled=12,decor.size=.35w}{v}
    \fmf{photon,width=0.7,tension=1.5}{v,o1}
  \marrow{ea}{down}{top}{}{i1,v}
  \marrow{eo}{up}{top}{}{v,i2}
  \end{fmfgraph*}}}
\stackrel{}{\underrightarrow{q}}
\parbox{12pt}{
  \begin{fmfgraph*}(23,30)
    \fmfstraight
    \fmfright{i1,i2}
    \fmfleft{o1}
    \fmf{plain,width=0.7,label=$p_3$,foreground=(1,,0.1,,0.1)}{v,i1}
    \fmf{plain,width=0.7,foreground=(1,,0.1,,0.1)}{v,i2}
    \fmfv{decor.shape=circle,decor.filled=12,decor.size=.35w}{v}
    \fmf{photon,width=0.7,tension=1.5}{v,o1}
    \marrow{ea}{down}{top}{}{i1,v}
    \marrow{eb}{up}{top}{}{v,i2}
  \end{fmfgraph*}}\hspace{0.3cm};\,\,\langle M_{5}\rangle=
  \hspace{0.08cm}
  \parbox{22pt}{
  \begin{fmfgraph*}(23,30)
    \fmfleft{i2,i1}
    \fmfright{o2,o1}
    \fmftop{t}
    \fmf{phantom,tension=0.2}{i1,v1,i2}
    \fmf{phantom}{o1,v2,o2}
    \fmf{phantom,tension=0.3}{v1,v2}
    \fmffreeze
    \fmf{plain,width=0.7
    ,foreground=(0.035,,0.168,,0.623)}{g,i1}
    \fmf{plain,width=0.7,tension=2.8}{g,v1}
    \fmf{plain,width=0.7,label=$p_1$,foreground=(0.035,,0.168,,0.623)}{i2,v1}
    \fmf{photon,width=0.7,tension=0}{t,v1}
    \fmf{photon,width=0.7}{v1,v2}
    \fmfv{decor.shape=circle,decor.filled=12,decor.size=.35w}{v1}
    \marrow{eb}{right}{rt}{$k$}{v1,t}
    \marrow{ea}{up}{top}{}{g,i1}
    \marrow{ec}{down}{bot}{}{i2,v1}
  \end{fmfgraph*}}
   \stackrel{}{\underleftarrow{q_3}}
\parbox{10 pt}{
  \begin{fmfgraph*}(23,30)
    \fmfstraight
    \fmfright{i1,i2}
    \fmfleft{o1}
    \fmf{plain,width=0.7,label=$p_3$,foreground=(1,,0.1,,0.1)}{v,i1}
    \fmf{plain,width=0.7
    ,foreground=(1,,0.1,,0.1)}{i2,v}
    \fmfv{decor.shape=circle,decor.filled=12,decor.size=.35w}{v}
    \fmf{photon,width=0.7,tension=1.5}{v,o1}
    \marrow{ea}{down}{top}{}{i1,v}
    \marrow{eb}{up}{bot}{}{v,i2}
  \end{fmfgraph*}}\hspace{0.5cm}
  +\,\,(1\leftrightarrow 3)\, .
\end{fmffile}
 \label{cuts m4 m5}
\end{equation}
\normalsize
In this work we denote by $A_{n}^{h,s}$ the transition amplitudes of a massive spin-$s$ state 
emitting $n-2$ massless particles. The case $h=1$, i.e. photon emission, has a long history in QED,  see for instance \cite{osti_4073049,book}. We start by reconsidering these objects in light of recent developments and unveil several new features.  As an introductory example, one can study
the soft expansion and double copy of $A^h_{3}$ and $A^h_{4}$
for a scalar source. It was shown in \cite{Bjerrum-Bohr:2013bxa}  via direct
computation that the double copy is realized
in a massive version of the KLT formula \cite{1986NuPhB.2691K,Bern:2008qj}: 
\begin{equation}
A_{n}^{{\rm ph},0}\times A_{n}^{{\rm ph},0}=K_{n}A_{n}^{{\rm gr},0}\label{eq:scklt},\quad n=3,4\,.
\end{equation}
with $K_{3}{=}1$ and $K_{4}{=}\frac{1}{2}\frac{k_1{\cdot}k_2}{p_{1}{\cdot}k_{1}\,p_{1}{\cdot}k_{2}}$ \footnote{We restore units in the final results and redefine $-iJ_{\rm CS}\rightarrow J_{\rm here}$ with respect to \cite{Cachazo:2014fwa}. We work in mostly minus signature.}, where $p_1{+}k_1{=}p_2{+}k_2$ and $k_i$ is massless.
While $A_{3}$ corresponds to a classical on-shell current and can be
used to evaluate conservative effects, it is not enough for the
computation of radiative effects even at LO in the coupling \cite{Shen:2018ebu,Goldberger:2017ogt}.
This can be understood from the fact that it does not posses orbit multipoles,
in contrast with $A_{4}$. Let us define orbit multipoles as each
of the terms appearing in the soft-expansion of $A_n$ with respect to an external
photon/graviton. Such expansion is trivial for $A_{3}$. For $A_{4}$, it
truncates at subleading order for photons \cite{PhysRev.96.1428,PhysRev.96.1433}. 
It follows
from \eqref{eq:scklt} that it truncates at subsubleading order for
gravitons. As a consequence, both amplitudes can be directly
constructed via Soft Theorems without the need for a Lagrangian.
The only seed is the amplitude $A_{3}^{h}(p_{1},k_1)=( \epsilon{\cdot}p_{1})^{h}$
which is fixed up to a constant using 3-pt. kinematics. Let us then write the soft expansion
with respect to $k_{2}\rightarrow0$ as
\begin{equation}
A_{4}^{{\rm ph}}=\frac{1}{2}\sum_{a{=}1,2}\frac{\epsilon_{2}{\cdot}p_{a}}{k_{2}{\cdot}p_{a}}e^{\frac{2F_{2}{\cdot} J_{a}}{\epsilon_{2}{\cdot} p_{a}}}A_{3}^{{\rm ph}}{=}\frac{1}{2}\left[\frac{p_{1}{\cdot}\epsilon_{1}F_{k}}{p_{1}{\cdot}k_{2}\,p_{2}{\cdot}k_{2}}{-}\frac{F_{\epsilon}}{p_{1}{\cdot}k_{2}}
\right]\label{eq:scphcompton},
\end{equation}
where $F_{2}{\cdot} J_{a}=F_{2}^{\mu\nu}J_{a\mu\nu}$ is the action
of the angular momentum operator \cite{Cachazo:2014fwa} on the corresponding particle and
$F_{k}=p_{1}{\cdot}F_{2}{\cdot}k_{1}$, $F_{\epsilon}=p_{1}{\cdot}F_{2}{\cdot}\epsilon_{1}$.
Analogously
\begin{equation}
    \begin{split}
      A_{4}^{{\rm gr}} & {=}  \sum_{p_a{=}p_{1},p_{2},k_{1}}\frac{1}{2}\frac{(\epsilon_{2}\cdot p_{a})^{2}}{k_{2}\cdot p_{a}}e^{\frac{2F_{2}\cdot J_{a}}{\epsilon_{2}\cdot p_{a}}}A_{3}^{{\rm gr}}=\frac{1}{2k_1{\cdot}k_2}\times \\
   &\quad \left[\frac{(p_{1}{\cdot}\epsilon_{1})^{2}}{p_{1}{\cdot}k_{2}p_{2}{\cdot}k_{2}}F_{k}^{2}-2\frac{p_{1}{\cdot}\epsilon_{1}}{p_{1}{\cdot}k_{2}}F_{k}F_{\epsilon}+\frac{p_{2}{\cdot}k_{2}}{p_{1}{\cdot}k_{2}}F_{\epsilon}^{2}\right]\label{eq:scgrcompton}.  
    \end{split}
\end{equation}
Given that $F_{2}{\cdot} J_{a}$ truncates when acting on $A_{3}$,
the exponential has been inserted to get the soft-expansion at the
desired order. The result not only manifests the double copy \eqref{eq:scklt}
but, as we will show, it generates the frequency expansion  of classical radiation
in these theories. The first term of the soft expansion therefore determines the
dipole radiation formula in EM and the Einstein's quadrupole radiation
in GR, whereas the subleading orders contribute to electric/magnetic higher multipoles \cite{Jackson:100964}. For bodies with long range interactions as in \eqref{cuts m4 m5}, this expands in powers of their orbital angular momentum, hence the name orbit multipole.

As a final remark note that we have written $\epsilon_{\mu\nu}=\epsilon_{\mu}\epsilon_{\nu}$
for the graviton polarization. This drastically simplifies the notation
and trivially projects-out Dilaton and Kalb-Ramond fields arising
in the double copy of $A_{n}$ amplitudes, in great contrast with
the cases of $M_{4}$ and $M_{5}$ \cite{Luna:2017dtq}. Although this only covers
$D-2$ states, for $D>4$ the remaining ones are obtained
by setting $\epsilon_{\mu}\epsilon_{\nu}\to\epsilon_{\mu\nu}^{{\rm TT}}$ (transverse-traceless tensor)
in our results.
\subsection*{Spin-Multipoles}
Our goal is to promote the above discussion for the case of spinning
sources, which introduces a rich new set of structures. In fact, the
seed $A_{3}^{h,s}$ is not unique and contains a soft expansion
encoding corrections to $A_{3}^{h,0}$ \cite{osti_4073049,PhysRev.96.1428,PhysRev.110.974,Guevara:2018wpp}.  
As the spin is the only quantum number available for the massive state, for any $n$ we can write \footnote{Formally, this can be argued via the generalized Wigner-Eckart theorem of e.g. \cite{Agrawala1980}, even if the group is non-compact.}
 \begin{equation}
A_{n}^{h,s}(J)={\rm \mathcal{H}}_n\times\sum_{j=0}^{\infty}\omega_{\mu_{1}\cdots\mu_{2j}}^{(2j)}J^{\mu_{1}\mu_{2}}\cdots J^{\mu_{2j-1}\mu_{2j}}\label{eq:multiexp},
\end{equation}
where $J^{\mu\nu}$ acts on spin-$s$ states. Products of $J^{\mu\nu}$ are
symmetrized since $[J,J]\sim J$ can be put in terms of lower multipoles. The sum is then guaranteed to truncate due to the Cayley-Hamilton theorem.
For $n=3$ we encode the helicity of the photon/graviton in the prefactor
$\mathcal{H}_3$.

To begin, let us consider photon emission for $s\in\{\frac{1}{2},1\}$ and
define its double copy. From two multipole operators $X$ and $X'$
acting on spin-$s$ states, we introduce an operator $X\odot X'$ 
acting on spin-$2s$ as
\begin{equation}
X\odot X'=\left\{ \begin{matrix}2^{-\left\lfloor D/2\right\rfloor }{\rm tr}(X\slashed{\varepsilon}_{1}\bar{X}'\slashed{\varepsilon}_{2})\,,\quad2s=1,\\
\phi_{1\mu_{1}\nu_{1}}\left(X_{\,\mu_{2}}^{\mu_{1}}X'{}_{\,\nu_{2}}^{\nu_{1}}\right)\phi_{2}^{\mu_{2}\nu_{2}}\,,\quad2s=2,
\end{matrix}\right.\label{eq:starprod}
\end{equation}
where $\varepsilon$ and $\phi$ are the respective massive polarizations
and $\bar{X}$ denotes charge conjugation. We will show that these operations can be
used to obtain scattering amplitudes in a gravity theory of
a massive spin-$2s$ field \cite{Bautista-Guevara}. Here we will only need the following extension of \eqref{eq:scklt}: 
\begin{equation}
A_{n}^{{\rm ph},s}\odot A_{n}^{{\rm ph},\tilde{s}}=K_{n}A_{n}^{{\rm gr},s+\tilde{s}}\,,\quad n=3,4\,.\label{eq:doublecopyspin}
\end{equation}
The case $s=0$, $\tilde{s}\neq0$ was introduced by Holstein et
al. \cite{Holstein:2006pq,Bjerrum-Bohr:2013bxa}. It was used 
to argue that the gyromagnetic ratios
of both $A_{n}^{{\rm ph},1}$ and $A_{n}^{{\rm gr},1}$ must coincide, setting $g=2$ as a natural value \cite{Holstein:2006pq,Chung:2018kqs}.
We introduce the case $s,\tilde{s}\neq0$ as a further universality
condition, and find it imposes strong restrictions on $A_{n}^{h,s}$
for higher spins. More importantly, it can be used to directly
obtain multipoles in the classical gravitational theory.

For \eqref{eq:doublecopyspin} to hold we need to put $A_{n}^{h,s}$
into the form \eqref{eq:multiexp} (although we will lift this restriction in \cite{Bautista-Guevara}). The coefficients $\omega^{(2j)}$ are universal
once we consider minimal-coupling amplitudes, which are obtained from
QED at $s=\frac{1}{2}$ and from the $W^{\pm}$-boson model at $s=1$ \cite{Holstein:2006pq}. The 3-pt. seeds in any dimension can be put as
\begin{equation}
A_{3}^{s,{\rm ph}}=\epsilon\cdot p_{1}\left(\mathbb{I}+J\right)\,,\quad J=\frac{\epsilon_{\mu}q_{\nu}J^{\mu\nu}}{\epsilon\cdot p_{1}}\,,\label{eq:3ptshalf}
\end{equation}
for $q=p_1{-}p_2$. Denoting  each operator by the corresponding ${\rm SO}(D-1,1)$ Young
diagram, i.e. $1=\mathbb{I}$ and $\ytableausetup{mathmode,boxsize=0.5em}\ydiagram{1,1}=J^{\mu\nu}$,
the operation \eqref{eq:starprod}  gives the rules
\begin{equation}
1_{s}\odot1_{s}=1_{2s}\,,\quad1_{s}\odot\ytableausetup{mathmode,boxsize=0.8em}\ydiagram{1,1}_{s}=\frac{1}{2}\ydiagram{1,1}_{2s}\,,\label{eq:rules}
\end{equation}
\begin{equation}
\ytableausetup{mathmode,boxsize=0.8em}\ydiagram{1,1}_{s}\odot\ydiagram{1,1}_{s}=\ydiagram{2,2}_{2s}+\ydiagram{2}_{2s}+\hat{1}_{2s}\, ,\label{eq:dcweyls2}
\end{equation}
which are a subset of the irreducible representations allowed
by the Clebsch-Gordan decomposition. Rule \eqref{eq:dcweyls2} is
explained in \eqref{eq:dc12} below. The first term we denote by $\Sigma^{\mu\nu\rho\sigma}$
and has the symmetries of a Weyl tensor, i.e. is the traceless part
of $\{J^{\mu\nu},J^{\rho\sigma}\}$. For instance, the $s=2$ amplitude
as obtained from \eqref{eq:doublecopyspin} is
\begin{equation}
A_{3}^{{\rm gr},2}=\left(\epsilon{\cdot}p_{1}\right)^{2}\phi_{2}{\cdot}\left(\mathbb{I}{+}\frac{\epsilon_{\mu}q_{\nu}J^{\mu\nu}}{\epsilon{\cdot} p_{1}}{+}\frac{W_{\mu\nu\alpha\beta}}{4(\epsilon{\cdot} p_{1})^{2}}\Sigma^{\mu\nu\alpha\beta}\right){\cdot}\phi_{1}\, ,\label{eq:3pts2}
\end{equation}
where $W_{\mu\nu\alpha\beta}:=q_{[\mu}\epsilon_{\nu]}q_{[\alpha}\epsilon_{\beta]}$
is the Weyl tensor of the graviton, 
reproducing the
expected Weyl-quadrupole coupling \cite{Goldberger:2004jt,Porto:2005ac,Porto:2006bt,Levi:2015msa,Chung:2018kqs}, 
 as shown in Appendix A.

To deeper understand these results, let us demand $A_{3}^{{\rm gr},s}$
to be constructible from the double copy \eqref{eq:doublecopyspin}
for \textit{any} spin:
\begin{equation}
  A_{3}^{{\rm gr,}s+\tilde{s}}(J^{\mu\nu}\oplus\tilde{J}^{\mu\nu})=A_{3}^{{\rm ph},s}(J^{\mu\nu})\odot A_{3}^{{\rm ph},\tilde{s}}(\tilde{J}^{\mu\nu})\, ,\label{eq:group}
\end{equation}
where $J^{\mu\nu}\oplus\tilde{J}^{\mu\nu}$ is the generator acting on a
spin $s+\tilde{s}$ representation. This relation yields the condition $A_{3}^{1,s}A_{3}^{1,\tilde{s}}=A_{3}^{1,s+\tilde{s}}A_{3}^{1,0}$ on the $J^{\mu\nu}$ operators. Using that $[J,\tilde{J}]=0$ and assuming the coefficients in \eqref{eq:multiexp} to be independent of the spin leads to
\begin{equation}
A_{3}^{h,s}(J)=\left({\rm \epsilon\cdot}p_{1}\right)^{h}\times e^{\omega_{\mu\nu}J^{\mu\nu}}\,,\quad h=1,2\label{eq:3ptexp}
\end{equation}
with $\omega_{\mu\nu}=\frac{k_{[\mu}\epsilon_{\nu]}}{\epsilon\cdot p_{1}}$ and $\mathcal{H}_3=\left(\epsilon{\cdot}p_{1}\right)^h$
 fixed by the previous examples. This easily recovers such cases
and matches the Lagrangian derivation \cite{Vaidya:2014kza} 
for $s\in\{\frac{1}{2},1,2\}$
in any dimension $D$. After some algebra, \eqref{eq:3ptexp} leads to the $D{=}4$
photon-current derived in \cite{Lorce:2009br,Lorce:2009bs} 
for \textit{arbitrary}
spin via completely different arguments. On the gravity side, it 
matches the Kerr stress-energy tensor derived in \cite{Vines:2017hyw} together with its spinor-helicity form recently found in \cite{Guevara:2018wpp}, as we show in Appendix B. For $s>h$ and $D>4$, \eqref{eq:3ptexp} contains
a pole in $\epsilon{\cdot}p$ which reflects such interactions being
non elementary \cite{Arkani-Hamed:2017jhn}. In Appendix A we show such pole cancels
for the classical multipoles and provide a local form of \eqref{eq:3ptexp}.

What is the meaning of the exponential $e^{J}$? It corresponds to
a finite Lorentz transformation induced by the massless emission.
That is, $p_{2}=e^{J}p_{1}$, hence for generic spin it maps the
state $|p_{1},\varepsilon_{1}\rangle$ into $|p_{2},\tilde{\varepsilon}_{2}\rangle$,
where $\tilde{\varepsilon}_{2}\neq\varepsilon_{2}$ is another polarization
for $p_{2}$. This means $e^{J}$ is composed both of a boost and
a ${\rm SO}(D-1)$ Wigner rotation. The boost can be removed in order
to match ${\rm SO}(D-1)$ multipoles in the classical theory, see
Appendix A. Also, as $e^{J}$ is a Lorentz transformation, $|\varepsilon_{2}\rangle$
must live in the same irrep as $|\varepsilon_{1}\rangle$. This means
that a projector is not needed when these objects are glued. A corollary
of this is a simple formula for the full factorizations of $A_{n}^{h,s}$,
e.g.
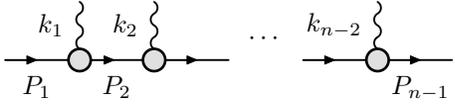
\begin{equation*}
     \begin{fmffile}{box}
    \begin{fmfgraph*}(211.3,22)
    \fmfleft{i1,i2}
    \fmfright{o1,o2}
    \fmf{phantom}{i1,v1,v2,v3,v4,v5,o1}
    \fmf{plain,width=0.7,label=$P_1$}{i1,v1}
     \fmf{plain,width=0.7,label=$P_2$}{v1,v2}
    \fmf{plain,width=0.7}{v2,v3}
    \fmf{phantom,label=$\cdots$}{v4,v3}
    \fmf{plain,width=0.7}{v4,v5}
    \fmf{plain,width=0.7,label=$P_{n-1}$}{v5,o1}
    \fmf{phantom}{o2,v6,v7,v8,v9,v10,i2}
    \fmf{photon,width=0.7,tension=0,label=$k_1$}{v1,v10}
    \fmf{photon,width=0.7,tension=0,label=$k_2$}{v9,v2}
    \fmf{phantom,tension=0}{v3,v8}
    \fmf{phantom,tension=0}{v4,v7}
    \fmf{photon,width=0.7,tension=0,label=$k_{n-2}$}{v6,v5}
    \fmfv{decor.shape=circle,decor.filled=12,decor.size=.04w}{v1}
    \fmfv{decor.shape=circle,decor.filled=12,decor.size=.04w}{v2}
    \fmfv{decor.shape=circle,decor.filled=12,decor.size=.04w}{v5}
    \marrow{a}{left}{top}{}{i1,v1}
    \marrow{b}{left}{top}{}{v1,v2}
    \marrow{d}{left}{top}{}{v2,v3}
    \marrow{f}{left}{top}{}{v4,v5}
    \marrow{c}{left}{top}{}{v5,o1}
    \end{fmfgraph*}
    \end{fmffile}\quad\quad \quad\quad   
\end{equation*}
\begin{equation}
   =\prod_{i}(P_{i}{\cdot}\epsilon_{i})^{h}\langle\varepsilon_{2}|e^{J_{n-2}}{\cdots}e^{J_{1}}|\varepsilon_{1}\rangle=\prod_{i}(P_{i}{\cdot}\epsilon_{i})^{h}\langle\varepsilon_{2}|\tilde{\varepsilon}_{2}\rangle,\label{eq:fact}\quad \quad
\end{equation}
where $P_{i}=p_{1}+k_{1}+\ldots+k_{i-1}$ and $J_{i}=\frac{k_{i\mu}\epsilon_{i\nu}J^{\mu\nu}}{\epsilon_{i}\cdot P_{i}}$.
Each 3-pt. amplitude here maps $P_{i}$ to $P_{i+1}$ and their composition
maps $p_{1}$ to $p_{2}$. The state $|\tilde{\varepsilon}_{2}\rangle$
depends on all $\{k_{i},\epsilon_{i}\}_{i=1}^{n}$ as well as their
ordering. This factorization is enough to obtain the classical spin-multipoles
of $M_{5}$ at least up to the quadrupole order we are interested
in. To see this, we use the Baker-Campbell-Hausdorff formula in \eqref{eq:fact} and get the form
\begin{equation}
    \begin{split}
        A_{4}^{{\rm ph},s}{=}& \frac{1}{2}\left[ \frac{p_{1}{\cdot}\epsilon_{1}p_{2}{\cdot}\epsilon_{2}}{p_{1}{\cdot} k_{1}}\langle\varepsilon_{2}|e^{J_{1}+J_{2}-\frac{1}{2}[J_{1},J_{2}]+\ldots}|\varepsilon_{1}\rangle{+}\label{eq:sumofexp}\right.\\
        & \left. 
        \frac{p_{2}{\cdot}\epsilon_{1}p_{1}{\cdot}\epsilon_{2}}{p_{2}{\cdot} k_{1}}\langle\varepsilon_{2}|e^{J'_{1}+J'_{2}+\frac{1}{2}[J'_{1},J'_{2}]+\ldots}|\varepsilon_{1}\rangle{+}{\rm c.t.}\right].
    \end{split}
\end{equation}
This is the spin analog of \eqref{eq:scphcompton}, where the exponential tracks the desired order. Setting $\mathcal{H}_4{=}\frac{1}{2}\frac{1}{p_1{\cdot}k_1 p_1{\cdot}k_2}$ in \eqref{eq:multiexp}, this gives for $s{\leq}1$
\begin{align}
\omega^{\mu\nu}_{(2)}{=} & \frac{p_{1}{\cdot}F_{1}{\cdot}p_{2}}{2}F_{2}^{\mu\nu}{+}\frac{p_{1}{\cdot}F_{2}{\cdot}p_{2}}{2}F_{1}^{\mu\nu}{+}\frac{p_{1}{\cdot}(k_{1}{+}k_{2})}{4}[F_{1}{,}F_{2}]^{\mu\nu}\nonumber \\
\omega^{\mu\nu\rho\sigma}_{(4)}{=} & \frac{k_{1}{\cdot}k_{2}}{16}\left(F_{1}^{\mu\nu}F_{2}^{\rho\sigma}+F_{2}^{\mu\nu}F_{1}^{\rho\sigma}\right)\label{eq:w2}
\end{align}
The role of the contact term in \eqref{eq:sumofexp} is to restore
gauge invariance. Here it is only needed for $\omega^{(0)}$, thus
by comparison  with \eqref{eq:scphcompton} one finds ${\rm c.t.}=\epsilon_{1}{\cdot}\epsilon_{2}$ and $\omega^{(0)}=p_1{\cdot}F_1 {\cdot}F_2{\cdot}p_1$.
Already for spin-$\frac{1}{2}$ it is clear that this decomposition
of the Compton amplitude is not evident at all from a Feynman-diagram
computation \cite{Bjerrum-Bohr:2013bxa,Ochirov:2018uyq}, 
whereas here it is direct.
A key point of this splitting is that under the \textit{double}
soft deformation $k_{3}=\tau\hat{k}_{3},k_{4}=\tau\hat{k}_{4}$, the
multipole $\omega^{(2j)}$ is $\mathcal{O}(\tau^j)$, whose
leading order will be the classical contribution. It is now instructive
to further decompose $A^{{\rm ph,s}}$ into irreps., which follows from
\begin{equation*}
\omega_{\mu\nu\rho\sigma}^{(4)}\{J^{\mu\nu},J^{\rho\sigma}\}=\left\{ \begin{matrix}\hat{1}[\omega^{(4)}]+[\omega^{(4)}]_{\mu\nu}Q^{\mu\nu},\quad s=1,\\
\hat{1}[\omega^{(4)}]+[\omega^{(4)}]_{\mu\nu\rho\sigma}\ell^{\mu\nu\rho\sigma},\,s=\frac{1}{2},
\end{matrix}\right.
\end{equation*}
where $\ell^{\mu\nu\rho\sigma}=\{J^{[\mu\nu},J^{\rho\sigma]}\}=\parbox{4pt}{\ytableausetup{mathmode,boxsize=0.2em}\ydiagram{1,1,1,1}},$
and
\begin{equation}
\hat{1}=\frac{J_{\mu\nu}J^{\mu\nu}}{2},\,\,Q^{\mu\nu}=\ytableausetup{mathmode,boxsize=0.4em}\ydiagram{2}=\{J^{\mu\rho},J_{\rho}^{\,\,\nu}\}+\frac{4}{D}\eta^{\mu\nu}\hat{1}.\label{eq:quadandl}
\end{equation}
The notation $[\omega^{(4)}]$ denotes the corresponding projections. Among them
we will be interested in the quadrupole operator $Q^{\mu\nu}$, only
present for $s\geq1$. 

Finally, $A_{4}^{{\rm gr,s}}$ is found from
\eqref{eq:doublecopyspin} and matches the Lagrangian result for $s\leq2$.
We have used that \eqref{eq:dcweyls2} reads
\begin{equation}
\begin{split}
      J_{s}^{\mu\nu}\odot J_{s}^{\rho\sigma} &{=}  \frac{1}{4}\Sigma_{2s}^{\mu\nu\rho\sigma}{+}\frac{\alpha_{D}}{D{-}2}\eta^{[\sigma[\nu}Q_{2s}^{\mu]\rho]}\\
  &\quad{+}\frac{\beta_{D}}{2D(D{-}1)}\eta^{\sigma[\nu}\eta^{\mu]\rho}\hat{1}_{2s} \,.
\end{split}
\label{eq:dc12}
\end{equation}
The normalizations $\alpha_{D},\beta_{D}$ depend solely on $D$.
However, it cancels out in the full computation  and hence we set $\alpha_{D}{=}\beta_{D}{=}1$
hereafter. Similarly, the condition $A_{4}^{{\rm ph},\frac{1}{2}}A_{4}^{{\rm ph},\frac{1}{2}}{=}A_{4}^{{\rm ph},0}A_{4}^{{\rm ph},1}$,
as implied by \eqref{eq:doublecopyspin}, can be traced at this order
to $[\omega^{(2)}\omega^{(2)}]_{\mu\nu}=[\omega^{(4)}]_{\mu\nu}\omega^{(0)}$,
which holds up to terms subleading in the double soft limit.
\subsection*{Classical Applications}
Very recently, Kosower et al. \cite{Kosower:2018adc} have provided a QFT derivation of
the following formulae
\begin{equation}
\Delta p^{\mu}{=}\int \frac{d^{D}q}{(2\pi)^{D-2} }\delta(2q{\cdot}p_{1})\delta(2q{\cdot}p_{3})q^{\mu}e^{iq\cdot b}\langle M^h_{4}\rangle \, ,\label{eq:deltap}
\end{equation}
\begin{equation}
\mathcal{J}_h(k){=}\int \frac{d^{D}q_1}{(2\pi)^{D-2} }\delta(2\,q_{1}{\cdot}p_{1})\delta(2\,q_{3}{\cdot}p_{3}) e^{iq_{1}b_{1}}e^{iq_{3}b_{3}}\langle M^h_{5}\rangle,\label{eq:JK}
\end{equation}
encoding classical observables at LO in the coupling \cite{Goldberger:2016iau}. Here $\Delta p^{\mu}=\frac{\partial\chi}{\partial b_{\mu}}$
is the (conservative) momentum deflection of a massive body in a classical
scattering setup, where the function $\chi$ is the scattering angle \cite{Bjerrum-Bohr:2018xdl}. The current $\mathcal{J}_h(k)$ reads $\epsilon_{\mu}J^{\mu}$
($h=1$) and $\epsilon_{\mu\nu}T^{\mu\nu}$ ($h=2$) and corresponds
to the field radiated at $r\rightarrow\infty$. Even though these were proven for $D=4$, matching with classical results shows that they hold in any $D$. As explained in \cite{Kosower:2018adc}, 
the classical limit $\langle M\rangle$ is obtained by rescaling $q_{i}=\hbar\hat{q}_{i}$
and $k=\hbar\hat{k}$, after which we can extract the leading order
in $\hbar$. We extend this rule to include spin by scaling
the angular-momentum as $J=\hbar^{-1}\hat{J}$, as in e.g. \cite{Guevara:2018wpp}. 

The $\hbar {\to} 0$ limit is captured by the cuts of $M_{4}$ and $M_{5}$ given in (\ref{cuts m4 m5}). For $M_4$, this was argued by one of the authors in \cite{Cachazo:2017jef}, where the classical piece was identified as the singularity in $q^2$ up to 1-loop, see also \cite{PhysRevD.38.3763}. For $M_5$, the key point is to introduce the average momentum transfer $q=\frac{q_1-q_3}{2}$, after which one expects the same construction to apply. In fact, noting that $d^D q_1=d^D q$ in \eqref{eq:JK} already shows that contact terms in $q^2$ appearing in $\langle M^h_5 \rangle $ will lead to local quantum contributions (details will be given somewhere else \cite{Bautista-Guevara}). 

To start with, consider $\langle M^h_{4}\rangle=\frac{n_h}{q^{2}}$ where $n_h$ a local
numerator. Its scalar parts are $n_{\rm ph}=p_{1}{\cdot} p_{3}$ and
\begin{equation}
\langle M_{4}^{{\rm gr}}\rangle=\frac{n_{\rm gr}}{q^{2}}=\frac{\sqrt{32\pi G}}{q^{2}}\left[(p_{1}{\cdot} p_{3})^{2}-\frac{m_{a}^{2}m_{b}^{2}}{D-2}\right]\label{eq:grnum}, 
\end{equation}
where the factor of $D-2$ arises from the graviton propagator. In
$D=4$ we can evaluate \eqref{eq:deltap} to recover the 1PM scattering
angle as in \cite{Bjerrum-Bohr:2018xdl}, 
first derived in the classical context
by Portilla \cite{Portilla_1979,Portilla_1980}. See below for spin effects. Moving to $\langle M^h_{5}\rangle$, the factorization of \eqref{cuts m4 m5} together with the classical limit imply the form
\begin{equation}
\langle M_{5}^{h}\rangle{=}\frac{1}{(q{\cdot}k)^{h-1}}\left[\frac{n_h^{(a)}}{(q^{2}{-}q{\cdot}k)(p_{1}{\cdot}k)^{2}}{\pm}\frac{n_h^{(b)}}{(q^{2}{+}q{\cdot}k)(p_{3}{\cdot}k)^{2}}\right]\,,\label{eq:newM5clas}
\end{equation}
where we pick $({-})$ for $h=2$. The spurious pole $q\cdot k$ arises from the $t$-channel of $A_{4}^{{\rm gr},s}$,
and its cancellation provides a nice check of our formula. This
further shows that the classical limits of $M_{4}$ and $M_{5}$ are
universal and do not depend on the spin
of the massive particles (nor the Lagrangian details if we assume $A^{h,s}_n$ are constructible). This was emphasized in \cite{Bjerrum-Bohr:2013bxa} 
at 4-pt. and is the first example of such universality at 5-pt.

\subsubsection*{Exponentiated Soft Theorem}
As an application of orbit multipoles let us study $\langle M_{5}^{{\rm gr}}\rangle$ for scalars. The numerators $n^{(a)}$ can be read off directly from \eqref{eq:scgrcompton}: Replacing $\epsilon_{1}$ by $p_{3}$, powers of the orbit multipole $F_{\epsilon}$
translate to powers of $F_{p}{=}p_{1}{\cdot}F{\cdot}p_{3}$, whereas $F_{k}$ now
becomes $F_{iq}{=}\eta_i (p_i{\cdot}F{\cdot}q)$, with $\eta_1{=}{-}1,\eta_3=1$. The soft expansion \eqref{eq:scgrcompton}
with respect to $k_{2}=k$ becomes
\begin{equation}
    n_{{\rm gr}}^{(a)}=\frac{F_{1q}^{2}}{2}e^{-\frac{F_{p}}{F_{1q}}(p_{1}{\cdot}k)\frac{\partial}{\partial(p_{1}{\cdot}p_{3})}}\left[(p_{1}{\cdot}p_{3})^{2}-\frac{m_{b}^{2}m_{a}^{2}}{D-2}\right].
\end{equation}
Further writing $\frac{1}{q^2\pm q\cdot k}=e^{\pm q\cdot k\frac{\partial}{\partial q^2}}\frac{1}{q^2}$
turns \eqref{eq:newM5clas} into
\begin{equation}
\boxed{
\langle M_{5}^{\rm gr}\rangle=\sum_{i=1,3}\mathcal{S}_{i}e^{\eta_i \left(F_{p}\frac{p_{i}{\cdot}k}{F_{iq}}\frac{\partial}{\partial(p_{1}{\cdot}p_{3})}+q{\cdot}k\frac{\partial}{\partial q^2}\right)}\langle M_{4}^{\rm gr}\rangle\,\label{eq:classicsoft}}
\end{equation}
where $\mathcal{S}_{i}{=}\frac{\eta_i}{2} \frac{F_{iq}^2}{(p_{i}{\cdot}k)^{2} q{\cdot}k}$ (for photons we find $\mathcal{S}_{i}{=}\frac{F_{iq}}{2(p_{i}{\cdot}k)^{2}}$). This expression
can be used to obtain $\langle M^{\rm gr}_{5}\rangle$ from $\langle M_{4}^{\rm gr}\rangle$
as an expansion in the graviton momenta $k^{\mu}$ to any desired
order (sub-subleading orders were studied in \cite{Laddha:2018myi,Sahoo:2018lxl,Ciafaloni:2018uwe}).
The spurious pole in $\mathcal{S}_{i}$ cancels out and one can check
explicitly that $\mathcal{S}_{1}+\mathcal{S}_{3}$ corresponds to
the $\hbar\to0$ limit of the Weinberg Soft Factor for the full $M_{5}$ \cite{Weinberg:1965nx}.
The first order of the exponential analogously corresponds to the
$\hbar\to0$ limit of the subleading soft factor of Low \cite{PhysRev.96.1428,PhysRev.110.974}.

Let us focus for simplicity on the leading order of \eqref{eq:classicsoft}. By considering bounded orbits with  $\omega \sim \frac{v}{r}$ the GW frequency expansion becomes a non-relativistic expansion \cite{Goldberger:2017vcg}.
It can be checked that the LO in fact leads to Einstein's Quadrupole Formula, see discussion below. For classical scattering we can use the LO to obtain the Memory Effect as $r{\to}\infty$. Plugging \eqref{eq:classicsoft}
into \eqref{eq:JK} we get
\begin{equation*}
    \int \frac{d^{D}q}{(2\pi )^{D-2}}\delta(2q\cdot p_{1})\delta(2q\cdot p_{3})e^{iq\cdot(b_{1}-b_{3})}\left(\sum_{i=1,3}\mathcal{S}_{i}\right)\langle M_{4}^{{\rm gr}}\rangle
\end{equation*}
as $k\to 0$. Evaluating the sum and using \eqref{eq:deltap} as a definition of
$\Delta p_{1}=-\Delta p_{3}$ we obtain
\begin{equation}
    \epsilon_{\mu\nu}T^{\mu\nu}=\frac{F_{p}/2}{p_{1}{\cdot}kp_{3}{\cdot}k}\left(\frac{p_{1}}{p_{1}{\cdot}k}{+}\frac{p_{3}}{p_{3}{\cdot}k}\right){\cdot}F{\cdot}\Delta p{+}\mathcal{O}(k^{0}),
\end{equation}
which at leading order in $\Delta p$ (or $G$, if restored) becomes
\begin{equation}
  T^{\mu\nu}(k)=\sqrt{8\pi G}\times\Delta\left[\frac{p_{1}^{\mu}p_{1}^{\nu}}{p_{1}{\cdot} k}+\frac{p_{3}^{\mu}p_{3}^{\nu}}{p_{3}{\cdot} k}\right]^{{\rm TT}}\, .
\end{equation}
In position space this gives
the burst memory wave derived by Braginsky and Thorne \cite{Braginsky1987} in $D=4$ (a $\frac{1}{4\pi r}$ factor arises from the ret. propagator as $r{\to}\infty$ \cite{Goldberger:2016iau,Hamada:2018cjj}), see also \cite{Pate:2017fgt,Mao:2017wvx,Satishchandran:2017pek} for $D>4$. Here we have provided a direct connection with the Soft Theorem \eqref{eq:classicsoft}, alternative to the expectation-value argument \cite{Strominger:2014pwa,Strominger:2017zoo}. This can also be seen as the Black Hole Bremsstrahlung of \cite{Luna:2016due,Luna:2016hge} generalized to consistently include the dynamics of the sources.
\subsubsection*{Classical Double Copy}
As the numerators in \cref{eq:grnum,eq:newM5clas}          
correspond
to $A_{n}^{h,s}$ amplitudes, the multipole double copy can be directly
promoted to $\langle M_{4}\rangle$ and $\langle M_{5}\rangle$. From
a classical perspective, the factorization of (\ref{cuts m4 m5}) implies that
the photon numerators can always be written as $n_{{\rm ph}}=t_{a\mu}t_{b}^{\mu}$
where $t_{a}$ and $t_{b}$ \textit{only} depend on particle 1 and
$3$ respectively. The simplest example is the scalar piece in $\langle M_{4}^{{\rm ph}}\rangle$,
where $t_{a}=p_{1}$ and $t_{b}=p_{3}$. The KLT formula \eqref{eq:doublecopyspin} translates
to
\begin{equation}
\boxed{
n_{{\rm gr}}=n_{{\rm ph}}\odot n_{{\rm ph}}-{\rm tr}(n_{{\rm ph}}\odot n_{{\rm ph}})\,\label{eq:numdc}}
\end{equation}
where we defined the trace operation as
${\rm tr}(n\odot n)=\frac{(t_{a\mu}\odot t_{a}^{\mu})(t_{b\mu}\odot t_{b}^{\mu})}{D-2}$. By combining \eqref{eq:numdc} with \cref{eq:grnum,eq:newM5clas}, 
this establishes for the first time a classical double-copy
formula that can be directly proved from the standard BCJ construction
\cite{Bautista-Guevara}. Moreover, up to this order it only requires as input Maxwell
radiation as opposed to gluon color-radiation \cite{Goldberger:2016iau,Goldberger:2017ogt} and contains no Dilaton/Axion states \cite{Johansson:2014zca,Luna:2017dtq,Goldberger:2017ogt}.

Let us start with $\langle M_{4}\rangle$ as example. To keep notation simple consider only particle $a$ to have spin. From \eqref{eq:3ptshalf}
we find that at the dipole level the numerator for $\langle M_{4}^{{\rm ph}}\rangle$
is $n_{\frac{1}{2}}^{{\rm ph}}=n_{{\rm 0}}^{{\rm ph}}+p_{3}{\cdot}J_{a}{\cdot}q$.
The gravity result follows from \eqref{eq:numdc} by dropping contact
terms in $q^2$. The rules \eqref{eq:rules} readily give the scalar and dipole
parts, including \eqref{eq:grnum}. For the quadrupole part, rule \eqref{eq:dc12}
gives
\begin{equation}
   \frac{(p_{3}{\cdot}J_{a}{\cdot}q)\odot(p_{3}{\cdot}J_{a}{\cdot}q)-{\rm tr(\cdots)}}{q^{2}}=\frac{1}{4}\frac{p_{3\mu}q_{\nu}p_{3\alpha}q_{\beta}\Sigma_a^{\mu\nu\alpha\beta}}{q^{2}}\,, 
\end{equation}
Using  \eqref{eq:weylproj}, the $\rm{SO}(D-1)$ quadrupole \cite{Porto:2005ac,Levi:2014gsa,Levi:2015msa} reads \footnote{Due to a transcription error in the first version of this preprint, the RHS of \eqref{weyl4} displayed the QED quadrupole term instead of the gravitational one.}
\begin{equation}
\frac{1}{4}\frac{p_{3\mu}q_{\nu}p_{3\alpha}q_{\beta}\Sigma_a^{\mu\nu\alpha\beta}}{q^{2}}\rightarrow\left((p_1{\cdot}p_3)^2{-}\frac{m^2_a m^2_b}{D{-}2}\right)\frac{q\cdot\bar{Q}_a\cdot q}{2(D{-}3)q^2 m_a^2}.\label{weyl4}
\end{equation}
Up to this order this agrees with the $D=4$ computation \cite{Vines:2017hyw,Guevara:2018wpp,o'connell-vines}. Agreement to all orders in spin is obtained from the formula
\eqref{eq:localmultip} in Appendix A. 

Moving to $\langle M_{5}\rangle$, in the examples that follow the
numerators $n_{{\rm ph}}$ can be read either from classical results
up to dipole order \cite{Luna:2017dtq,Kosower:2018adc,Goldberger:2017ogt,Li:2018qap}, from QED Bremsstrahlung,
or from \eqref{eq:scphcompton},\,\eqref{eq:3ptshalf} and \eqref{eq:w2}. They are all in agreement \footnote{Ref. \cite{Li:2018qap} may contain a typo. Reproducing the computation leads to a relative $({-})$ sign between eqs. 26 and 27.}. For photons, the scalar part is
\begin{equation}
    n_{0}^{(a)}{=}4e^3 p_1{\cdot}R_3{\cdot}F{\cdot}p_1, \quad 
    n_{0}^{(b)}{=}4e^3 p_3{\cdot}R_1{\cdot}F{\cdot}p_3, \label{nphsc}
\end{equation}
where $R_i^{\mu\nu}{=}p_{i}^{[\mu}(\eta_i 2q{-}k)^{\nu]}$.
For the spin part we have
\begin{align}
n_{\frac{1}{2}}^{(a)}{=} & n_{0}^{(a)}{-}2e^{3}\left[p_{1}{\cdot}R_{3}{\cdot}kF{\cdot}J_{a}{-}F_{1q}R_{3}{\cdot}J_{a}{+}p_{1}{\cdot}k\,[F,R_{3}]{\cdot}J_{a}\right],\nonumber \\
n_{\frac{1}{2}}^{(b)}{=} & n_{0}^{(b)}{+}2e^{3}p_{3}{\cdot}F{\cdot}\hat{R}_a{\cdot}p_{3},\label{eq:numerators 1/2}
\end{align}
with $\hat{R}^{\mu \nu}_a{=}\left(2q{+}k\right)^{[\mu }J_a^{\nu]\alpha}(2q{+k})_{\alpha}$. Recall these numerators live in the support of $\delta( p_i{\cdot}q_i)$ in \eqref{eq:JK}. Writing them as $n_{\frac{1}{2}}{=}t_{a}{\cdot}t_{b}$ one finds $t_{b}^{(a)}{=}p_{3}$
and $t_{a}^{(b)}=p_{1}{+}J_{a}{\cdot}(2q{+}k)$ as expected from their "3-pt. part". The scalar and dipole
pieces obtained from \eqref{eq:numdc} then recover the results of \cite{Luna:2017dtq,Goldberger:2017ogt,Li:2018qap} 
for Pure and Fat Gravity (we obtain the latter as the limit $D\to\infty$). This provides a strong cross-check of our method. Using \eqref{eq:dc12} we can also compute the quadrupole order. For instance, the $Q^{\mu\nu}$ piece reads
\begin{equation*}
     \begin{split}
          \frac{n^{(a)}|_Q}{q{\cdot}k}{=}&\frac{(32\pi G)^{\frac{3}{2}}}{8(D{-}2)}\biggl[\left(p_{1}{\cdot} p_{3}F_{1q}{-}p_{1}{\cdot}kF_{p}\right)\{R_3{,}F\}{\cdot}Q_{a}+ \hspace{1cm}\\
 &\frac{m_{b}^{2}}{(D{-}2)}\left(F_{1q}\{F{,}Y\}{\cdot}Q_{a}{-}2p_{1}{\cdot}k\,p_{1}{\cdot}F{\cdot}Q_{a}{\cdot}F{\cdot}q\right)\biggr],
     \end{split}
 \end{equation*}
with $Y^{\mu \nu}=p_1^{[\mu}(2q{-}k)^{\nu]}$, whereas $n^{(b)}|_Q=0$. As before, we have dropped contact terms in $q^2$ and used the support of $\delta(p_i{\cdot}q_i)$. This result  can be shown to agree with a much more lengthy computation of
the full $M_{5}^{{\rm gr}}$ using Feynman diagrams. At this order,
$M_{5}^{{\rm gr}}$ contains classical quadrupole pieces and quantum
scalar and dipole pieces. Interestingly, while the scalar part is trivial to identify, we have found that the dipole part can be cancelled by adding the spin-1 spin-0 interaction $(B_{\mu}\partial^{\mu}\phi)^{2}$ to the
Lagrangian, which signals its quantum nature.
\subsection*{Discussion}
We have shown that key techniques of Scattering Amplitudes such as soft
theorems and double copy can be promoted directly to study classical phenomena arising in Gravitational Waves (GW). These
techniques drastically streamline the computation of radiation and spin
effects; both are phenomenologically important for Black Holes, which are believed to be extremely spinning in nature \cite{Risaliti2013,Reis2014}. In that direction, one could for instance apply our formalism to derive the hexadecapole ($s=2$) order in radiation \cite{Marsat:2014xea,Siemonsen:2017yux} to LO in $G$ but all orders in $1/c$. We now outline some other directions:

\textit{The $A^{h,s}_n$ series}: Let us emphasize that these constitute building blocks even at loop orders \cite{Neill:2013wsa,Cachazo:2017jef,Bern:2019nnu}. For $s>2$ the amplitudes $A_4^{h,s}$ were studied in \cite{Chung:2018kqs}  in the context of the $\mathcal{O}(G^2)$ potential and were found to contain polynomial ambiguities. We expect our construction, including soft expansion and double copy, to be a criteria for resolving such ambiguities and lead to further classical predictions. In the scalar setup, we expect $A_n^{{\rm gr},0}$ to be relevant even for $n>4$. In fact, $A_5^{{\rm gr},0}$ as a double copy has been recently pointed out as a key ingredient in the computation of the $\mathcal{O}(G^3)$ potential by Bern et al. \cite{Bern:2019nnu}. All these results made strong use of the $D=4$ spinor-helicty formalism. Specializing our treatment of radiation to $D=4$ is also a natural future direction in the hunt of simplifications even at loop orders, as in \cite{Guevara:2017csg,Cachazo:2017jef}.

\textit{Soft Theorem/Memory Effect:} It would be interesting to understand the meaning of the higher orders of \eqref{eq:classicsoft}, considering for instance the Spin Memory Effect \cite{Pasterski:2015tva,Nichols:2017rqr}. Motivated by the infinite soft theorems of \cite{Hamada:2018vrw,Campiglia:2018dyi} one could expect the corrections are related to a hierarchy of symmetries. One may also incorporate spin contributions and study their interplay with such orders \cite{Hamada:2018cjj}. In the applications side, it is desirable to further investigate \eqref{eq:classicsoft} at loop level \cite{Bern:2014oka,He:2014bga}, which could lead to a simple way of obtaining $\langle M_5\rangle$ from $\langle M_4\rangle$.

\textit{Generic Orbits:}  
For orbits more general than scattering $\mathcal{J}(k)$ does not have the support of $\delta(2 p_i{\cdot}q_i)$ \cite{Goldberger:2017vcg,Shen:2018ebu}. In fact, for bounded orbits it contains the subleading terms $p_i{\cdot}q_i \sim \omega$. Very nicely, by keeping such terms in the classical calculation we have checked they match with eqs. \eqref{nphsc},\eqref{eq:numerators 1/2}, which in turn arise from the form in \eqref{eq:w2} via a natural "$F{\to}R$ replacement". One could then try to explore the gravity case by combining our results with the EFT treatment of bounded orbits and their EOMs \cite{Porto:2016pyg}.

Acknowledgements: We thank Freddy Cachazo, Sebastian Mizera, Alex Ochirov and Nils Siemonsen for useful discussions and comments on the draft. We especially thank Justin Vines for clarifications regarding asymptotic trajectories and spin phenomenology. Y.F.B. is funded by the  Allan Carswell scholarship of the faculty of science YU. A.G. thanks CONICYT for financial support. Research at Perimeter Institute is supported by the Government of Canada through Industry Canada and by the Province of Ontario through the Ministry of Research \& Innovation.
\subsection*{Appendix A: From $\rm{SO}(D-1,1)$ to $\rm{SO}(D-1)$ multipoles}

In order to compare
with classical results for spinning bodies it is sometimes necessary
to choose a frame through the Spin Supplementary
Condition (SSC). Let us show how this arises from our setup.

We have shown that the spin multipoles correspond to finite $\rm{SO}(D-1,1)$
transformations which map $p_{1}$ $\rightarrow$ $p_{2}$. Such Lorentz
transformations are composed of both a boost and a $\rm{SO}(D-1)$ Wigner
rotation. Spin multipoles of a massive spinning body are defined with
respect to a reference time-like direction and form irreps. of $\rm{SO}(D-1)$
acting on the transverse directions \cite{Levi:2015msa,Levi:2018nxp}. Hence, it is natural
to identify such action with Wigner rotations of the massive states
entering our amplitude. A simple choice for the time-like direction
is the average momentum $u=\frac{p}{m}=\frac{p_{1}+p_{2}}{2m}$. In
this frame boosts are obtained as $K^{\nu}=u_{\nu}J^{\mu\nu}$ whereas
Wigner rotations read $S^{\mu\nu}=J^{\mu\nu}-2u^{[\mu}K^{\nu]}$.
Adopting $S^{\mu\nu}$ as classical spin tensor then corresponds to
the \textit{covariant} SSC, i.e. $u_{\nu}S^{\nu\mu}=0$ \cite{Porto:2008tb,Porto:2008jj,Vines:2017hyw}
(other choice was used in \cite{Chung:2018kqs,Holstein:2008sx}).
The momenta
$p_{1}$ and $p_{2}$ can be aligned canonically to $p$ through the
boost,
\begin{equation}
p_{1}=e^{\frac{q}{2m}\cdot K}p\,,\quad p_{2}=e^{-\frac{q}{2m}\cdot K}p\,,\label{eq:boost}
\end{equation}
which defines canonical polarization vectors $\varepsilon$, $\tilde{\varepsilon}$
for $p$ through (recall $p_{2}$ is outgoing):
\begin{equation}
   \varepsilon_{1}{=}e^{\frac{q}{2m}\cdot K}\,\varepsilon\,\,,\quad\varepsilon_{2}{=}\tilde{\varepsilon}\,e^{\frac{q}{2m}\cdot K}\,.
\end{equation}
This replacement can then be applied to the multipole expansion \eqref{eq:multiexp},
yielding an extra power of $q$ for each power of $J$, hence preserving
the $\hbar$-scaling. We find
\begin{eqnarray}
\varepsilon_{1}{\cdot}\varepsilon_{2} & {=} &
\varepsilon{\cdot}\tilde{\varepsilon}{+}\frac{1}{m}q_{\mu}\varepsilon K^{\mu}\tilde{\varepsilon}{+}\mathcal{O}(K^{2})\,,\\
\varepsilon_{1}J^{\mu\nu}\varepsilon_{2} & {=}& 
\varepsilon S^{\mu\nu}\tilde{\varepsilon}{+}2u^{[\mu}\varepsilon K^{\nu]}\tilde{\varepsilon}{+}\nonumber \\
 & &
 \frac{q_{\alpha}}{m}\varepsilon\{K^{\alpha}{,}S^{\mu\nu}\}\tilde{\varepsilon}{+}\mathcal{O}(K^{2})\,,\\
\varepsilon_{1}\{J^{\mu\nu}{,}J^{\rho\sigma}\}\varepsilon_{2} &{ =}  & 
\varepsilon\{S^{\mu\nu}{,}S^{\rho\sigma}\}\tilde{\varepsilon}{+}\mathcal{O}(K)\,,\label{eq:quadK}
\end{eqnarray}
(for generic spin $K$ and $S$ are independent). In terms of irreducible
representations this decomposition can be thought of as branching
$\rm{SO}(D-1,1)$ into $\rm{SO}(D-1)$ \cite{Bekaert:2006py}. For instance, the dipole branches as
$\ytableausetup{mathmode,boxsize=0.6em}\ydiagram{1,1}\rightarrow\ydiagram{1,1}+\ydiagram{1}$,
which is a transverse dipole plus a transverse vector irrep, $K^{\mu}$.
In the same way, in general the $\ytableausetup{mathmode,boxsize=0.6em}\ydiagram{2,2}$
irrep. of $\rm{SO}(D-1,1)$ also contains a $\ytableausetup{mathmode,boxsize=0.6em}\ydiagram{2}$
piece for $\rm{SO}(D-1)$. This is the reason we can extract a quadrupole
from Weyl piece in \eqref{weyl4}, namely by combining \eqref{eq:quadK} with
the replacement rule
\begin{equation}
\{S^{\mu\nu},S^{\rho\sigma}\}=\frac{2}{D{-}3}\left(\bar{\eta}^{\sigma[\mu}\bar{Q}^{\nu]\rho}{-}\bar{\eta}^{\rho[\mu}\bar{Q}^{\nu]\sigma}\right)+{\rm other}\,\,{\rm irreps}\,\label{eq:SStoQ}
\end{equation}
where $\bar{\eta}^{\mu\nu}=\eta^{\mu\nu}-u^{\mu}u^{\nu}$. Thus we
have the identity (c.f. \cite{Steinhoff:2012rw,Chen:2019hac})
\begin{equation}
    \begin{split}
      \omega_{\mu\nu\rho\sigma}\Sigma^{\mu\nu\rho\sigma}  &=  [\omega]_{\mu\nu\rho\sigma}^{\ytableausetup{mathmode,boxsize=0.15em}\ydiagram{2,2}}\langle\varepsilon_{1}|\{J^{\mu\nu}{,}J^{\rho\sigma}\}|\varepsilon_{2}\rangle\label{eq:weylproj}\,,\\
  &=  \frac{4}{D-3}[\omega]_{\mu\nu\rho\sigma}^{\ytableausetup{mathmode,boxsize=0.15em}\ydiagram{2,2}}u^{\mu}\bar{Q}^{\nu\rho}u^{\sigma}+O(K)\,.
    \end{split}
\end{equation}
For instance, we extract a quadrupole contribution from $A_{3}^{h,s}$ in \eqref{eq:3pts2}:
\begin{equation}
A_{3}^{h,s}|_{\bar{Q}} { =}  \frac{1}{4}\left(\epsilon\cdot p_{1}\right)^{h}\frac{q\cdot\bar{Q}\cdot q}{D-3}\label{eq:quad3pt2}\,.
\end{equation}
Of course, the $\rm{SO}(D-1,1)$ quadrupole present in $A_{4}^{h,s}$ also
contains a $\rm{SO}(D-1)$ quadrupole. It follows from \eqref{eq:quadK}
that it can be read through
\begin{equation}
    Q^{\mu\sigma}{=}\bar{Q}^{\mu\sigma}{-}\frac{4}{D(D-1)}\bar{\eta}^{\mu\sigma}S^{2}{+}\mathcal{O}(K)\,.
\end{equation}
In general the $\rm{SO}(D-1)$ multipoles defined
through the covariant SSC are given directly from the $\rm{SO}(D-1,1)$ ones,
up to $O(K)$ terms. Due to unitarity, one expects the latter to drop from the amplitude, at least for $A_3$. Let us show explicitly how this happens. Note that 3-pt. kinematics implies $[q{\cdot} K,q{\cdot} J{\cdot}\epsilon]=0$
and hence the spin piece of the 3-pt. amplitude \eqref{eq:3ptexp} reads
\begin{equation}
    \begin{split}
        \varepsilon_{1}e^{\frac{q\cdot J\cdot\epsilon}{\epsilon\cdot p}}\varepsilon_{2} &{ = } \tilde{\varepsilon}\exp\left(\frac{q_{\mu}\epsilon_{\nu}J^{\mu\nu}}{\epsilon\cdot p}{+}\frac{q_{\mu}K^{\mu}}{m}\right)\varepsilon\,=\tilde{\varepsilon}e^{\mathcal{S}}\varepsilon\, \\
 & =  \sum_{n=0}^{\infty}\frac{1}{n!}\tilde{\varepsilon}\left(\frac{q_{\mu}\epsilon_{\nu}S^{\mu\nu}}{\epsilon\cdot p}\right)^{n}\varepsilon\label{eq:expred-2}\,,
    \end{split}
\end{equation}
where one can check that the sum truncates at order $2s$. Thus the boost \eqref{eq:boost} is effectively subtracted from the
finite Lorentz transformation leading to the interpretation of the
3-pt. formula as a little-group rotation induced via photon/graviton
emission. We end with a comment on the case $s>h$ and $D>4$: Note
that the pole $\epsilon\cdot p$ cancels in \eqref{eq:quad3pt2} for
any dimension. This means we can provide a local form of the 3-pt.
amplitude which contains the same multipoles as the exponential. For instance,
\begin{equation}
    \begin{split}
       \bar{A}_{3}^{{\rm ph},2}  {=}&  \left(\epsilon{\cdot} p\right)\phi_{2}{\cdot}\left(\mathbb{I}{+}\frac{\epsilon_{\mu}q_{\nu}J^{\mu\nu}}{\epsilon{\cdot} p}{+}\frac{q_{\mu}q_{\rho}}{4m^{2}\,\epsilon{\cdot} p}\times\right.\\
 &  \left.\left[\epsilon_{\nu}p_{\sigma}{+}\epsilon_{\sigma}p_{\nu}{-}\frac{\eta_{\nu\sigma}\left(\epsilon{\cdot} p\right)}{D{-}3}\right]\{J^{\mu\nu}{,}J^{\rho\sigma}\}\right){\cdot}\phi_{1} \,,
    \end{split}
\end{equation}
also yields \eqref{eq:quad3pt2} and reduces to \eqref{eq:3ptexp} in $D=4$. In general the $2^{n}$-poles
\cite{Levi:2018nxp,Vines:2017hyw} of \eqref{eq:expred-2} are obtained by performing $\left\lfloor \frac{n}{2}\right\rfloor $
traces with the spatial metric $\bar{\eta}^{\alpha\beta}$ appearing
in \eqref{eq:SStoQ}. The result takes the local form
\begin{equation}
    \begin{split}
        \left.A_{3}^{h,s}\right|_{2^{n}-{\rm poles}}  {=}&  \left(\epsilon{\cdot} p\right)^{h}\sum_{n=0}^{\infty}\left(\alpha_{n}{+}\beta_{n}\frac{q_{\mu}\epsilon_{\nu}S^{\mu\nu}}{\epsilon{\cdot} p}\right) \\
 &   \times\bar{Q}_{\mu_{1}{\ldots}\mu_{2n}}^{(n)}q^{\mu_{1}}{\cdots} q^{\mu_{2n}}\label{eq:localmultip}\,,
    \end{split}
\end{equation}
where $\alpha_{n}$, $\beta_{n}$ depend on the dimension $D$, and
$\bar{Q}{}^{(n)}$ are the transverse multipoles. In four dimensions
we find $\bar{Q}^{(n)}$ to be a tensor product of the Pauli-Lubanski
vector $S^{\mu}$ \cite{Levi:2018nxp,Chung:2018kqs},
and $\alpha_{n}=\frac{m^{-2n}}{(2n)!},$
$\beta_{n}=\frac{m^{-2n}}{(2n+1)!}$.
\subsection*{Appendix B: Spinor-Helicity Formulae}

Here we show the exponential forms presented here for spin-multipoles
contain as particular cases the ones of \cite{Guevara:2018wpp}, which implemented
massive spinor-helicity variables in $D=4$ \cite{Arkani-Hamed:2017jhn}. Consider
first $A_{3}^{{\rm gr},s}$: For plus helicity of the graviton, the
expression derived in \cite{Guevara:2018wpp} reads
\begin{equation}
A_{3,+}^{{\rm gr},s}=\frac{(p\cdot\epsilon)^{2}}{m^{2s}}\langle2|^{2s}e^{\frac{k_{\mu}\epsilon_{\nu}J^{\mu\nu}}{p\cdot\epsilon}}|1\rangle^{2s}\label{eq:4dsetup}\,,
\end{equation}
where $\epsilon{=}\epsilon^{+}$ carries the graviton helicity and $|\lambda\rangle^{2s}$
stands for the product $|\lambda^{(a_{1}}\rangle_{\alpha_{1}}\cdots|\lambda^{a_{2s})}\rangle_{\alpha_{2s}}$
of ${\rm SL}(2,\mathbb{C})$ spinors associated to each massive particle.
The generator $J^{\mu\nu}$ in the exponent acts on such chiral
representation. The labels $a_{i}$ are completely symmetrized little-group
indices. The explicit construction of the massive spinors is not needed
here (c.f. \cite{Arkani-Hamed:2017jhn}), but solely the fact that spin-$s$ polarization
tensors can be expressed compactly as
\begin{equation}
\varepsilon_{1} { =}  \frac{1}{m^{s}}|1\rangle^{s}|1]^{s}\,,\quad\varepsilon_{2}{=}\frac{1}{m^{s}}|2\rangle^{s}|2]^{s}\,,
\end{equation}
where $|1^{a}]_{\dot{\alpha}}$ and $|2^{a}]_{\dot{\alpha}}$ live
in the antichiral representation of ${\rm SL}(2,\mathbb{C})$. Inserting
them  into \eqref{eq:3ptexp} we obtain 
\begin{equation}
\langle\varepsilon_{2}|A_{3}^{{\rm gr},s}|\varepsilon_{1}\rangle{=}\frac{(p\cdot\epsilon)^{2}}{m^{2s}}\langle2|^{s}e^{\frac{k_{\mu}\epsilon_{\nu}J^{\mu\nu}}{p\cdot\epsilon}}|1\rangle^{s}[2|^{s}e^{\frac{k_{\mu}\epsilon_{\nu}\tilde{J}^{\mu\nu}}{p\cdot\epsilon}}|1]^{s}\,,\label{eq:3ptredux}
\end{equation}
where $J^{\mu\nu}$ and $\tilde{J}^{\mu\nu}$ are given by
\begin{eqnarray}
J^{\mu\nu} & {=} & \frac{1}{2}\mbox{\ensuremath{\sigma}}^{\mu\nu}\otimes\mathbb{I}^{\otimes(s-1)}{+}\mathbb{I}\otimes\frac{1}{2}\mbox{\ensuremath{\sigma}}^{\mu\nu}\otimes\mathbb{I}^{\otimes(s-2)}{+}{\cdots}\,,\quad\\
\tilde{J}^{\mu\nu} & {=} & \frac{1}{2}\mbox{\ensuremath{\tilde{\sigma}}}^{\mu\nu}\otimes\mathbb{I}^{\otimes(s-1)}{+}\mathbb{I}\otimes\frac{1}{2}\mbox{\ensuremath{\tilde{\sigma}}}^{\mu\nu}\otimes\mathbb{I}^{\otimes(s-2)}{+}{\cdots}\,,\quad
\end{eqnarray}
with $\sigma^{\mu\nu}=\sigma^{[\mu}\tilde{\sigma}^{\nu]}$
and $\tilde{\sigma}^{\mu\nu}=\tilde{\sigma}^{[\mu}\sigma^{\nu]}$.
They satisfy the self-duality conditions
\begin{equation}
    J^{\mu\nu} =  \frac{i}{2}\epsilon^{\mu\nu\rho\sigma}J_{\rho\sigma}\,\,,\quad
\tilde{J}^{\mu\nu} = -\frac{i}{2}\epsilon^{\mu\nu\rho\sigma}\tilde{J}_{\rho\sigma}\,.
\end{equation}
As it is well known, choosing the graviton to have plus helicity leads
to a self-dual field strength tensor, which in turn implies that $k_{[\mu}\epsilon_{\nu]}^{+}\tilde{J}^{\mu\nu}=0$.
Then \eqref{eq:3ptredux} reads

\begin{equation}
\langle\varepsilon_{2}|A_{3}^{{\rm gr},s}|\varepsilon_{1}\rangle=\frac{(p\cdot\epsilon)^{2}}{m^{2s}}\langle2|^{s}e^{\frac{k_{\mu}\epsilon_{\nu}J^{\mu\nu}}{p\cdot\epsilon}}|1\rangle^{s}[21]^{s}\,.
\end{equation}
We can now plug the identity
$
[21]^{s}{=}\langle2|^{s}e^{\frac{k_{\mu}\epsilon_{\nu}J^{\mu\nu}}{p\cdot\epsilon}}|1\rangle^{s}
$ from \cite{Guevara:2018wpp}
to obtain:
\begin{equation}
  \langle\varepsilon_{2}|A_{3}^{{\rm {\rm gr}},s}|\varepsilon_{1}\rangle {=}  \frac{(p{\cdot}\epsilon)^{2}}{m^{2s}}\langle2|^{s}e^{\frac{k_{\mu}\epsilon_{\nu}J^{\mu\nu}}{p\cdot\epsilon}}|1\rangle^{s} \langle2|^{s}e^{\frac{k_{\mu}\epsilon_{\nu}J^{\mu\nu}}{p\cdot\epsilon}}|1\rangle^{s}\label{eq:prematch} .
\end{equation}
which has the structure of our formula \eqref{eq:group}, now in "spinor space". Extending the generators $J^{\mu\nu}$ to act on $2s$ slots, i.e.
$J^{\mu\nu}\otimes\mathbb{I}^{s}+\mathbb{I}^{s}\otimes J^{\mu\nu}\rightarrow J^{\mu\nu}$,
then recovers \eqref{eq:4dsetup}. Consider now $A_{4,{+}{-}}^{{\rm gr},s}$
for $s\leq2$ as given in \cite{Guevara:2018wpp}, where $(+-)$ denotes the helicity
of the gravitons $k_{1}=|\hat{1}]\langle\hat{1}|$ and $k_{2}=|\hat{2}]\langle\hat{2}|$,

\begin{equation}
A_{4,++}^{{\rm gr},s}=\frac{\langle\hat{1}|P_{1}|\hat{2}]^{4}m^{-2s}}{p_1{\cdot}k_1\,p_1{\cdot}k_2\, k_1{\cdot}k_2}\langle2|^{2s}e^{\frac{k_{1\mu}\epsilon_{1\nu}J^{\mu\nu}}{p\cdot\epsilon_{1}}}|1\rangle^{2s}\label{eq:spinorcomp}\,.
\end{equation}
In order to match this we double copy our formula \eqref{eq:sumofexp}.
The sum in \eqref{eq:sumofexp} exponentiates if we impose $[J_{1},J_{2}]=0$,
which in turn is only possible if the polarizations are aligned,
i.e. $\epsilon_{1}\propto\epsilon_{2}$. When the states have opposite
helicity this can be achieved via a gauge choice. This yields
\begin{equation}
\frac{k_{1\mu}\epsilon_{1\nu}J^{\mu\nu}}{p_{1}\cdot\epsilon_{1}}+\frac{k_{2\mu}\epsilon_{2\nu}J^{\mu\nu}}{p_{2}\cdot\epsilon_{2}}=\frac{k_{\mu}\epsilon_{1\nu}J^{\mu\nu}}{p\cdot\epsilon_{1}},
\end{equation}
where $k=k_{1}+k_{2}$. Expression \eqref{eq:sumofexp} thus becomes
\begin{equation}
\left.A_{4}^{{\rm ph},s}\right|_{\epsilon_{1}\propto\epsilon_{2}}{=}\frac{p_{1}{\cdot}\epsilon_{1}\,p_{2}{\cdot}\epsilon_{2}\,k_1{\cdot}k_2}{p_{1}{\cdot}k_{1}p_{1}{\cdot}k_{2}}\langle\varepsilon_{1}|e^{\frac{k_{\mu}\epsilon_{1\nu}J^{\mu\nu}}{p{\cdot}\epsilon_{1}}}|\varepsilon_{2}\rangle\,.
\end{equation}
(note that ${\rm ct}=\epsilon_{1}\cdot\epsilon_{2}$ drops out). The
formula \eqref{eq:doublecopyspin} gives
\begin{equation}
\left.A_{4}^{{\rm gr},s}\right|_{\epsilon_{1}\propto\epsilon_{2}}=\frac{(p_{1}{\cdot}\epsilon_{1})^{2}(p_{2}{\cdot}\epsilon_{2})^{2}}{p_{1}{\cdot}k_{1}\,p_{1}{\cdot}k_{2}\,k_1{\cdot}k_2}\langle\varepsilon_{1}|e^{\frac{k_{\mu}\epsilon_{1\nu}J^{\mu\nu}}{p\cdot\epsilon_{1}}}|\varepsilon_{2}\rangle\label{eq:gr4dcomp}\,,
\end{equation}
for $s\leq2$. This can be shown to match \eqref{eq:spinorcomp} following
the same derivation as before and fixing  $\epsilon_{1}=\frac{|\hat{1}\rangle[\hat{2}|}{[\hat{1}\hat{2}]}$,
$\epsilon_{2}=\frac{|\hat{1}\rangle[\hat{2}|}{\langle\hat{1}\hat{2}\rangle}$.
Note finally that, even though in any dimension $D$ there is an helicity
choice such that \eqref{eq:sumofexp} becomes \eqref{eq:gr4dcomp},
the factorization of (\ref{cuts m4 m5}) requires to sum over all helicities of
internal gravitons.
\bibliography{references}
\bibliographystyle{aipnum4-1}
\end{document}

%% file: author_list.tex

\author{Yilber Fabian Bautista} \email[]{ybautistachivata@perimeterinstitute.ca} \affiliation{Perimeter Institute for Theoretical Physics, Waterloo, ON N2L 2Y5, Canada}\affiliation{ Department of Physics   and  Astronomy, York University, Toronto, Ontario, M3J 1P3, Canada.}
\author{Alfredo Guevara } \email[]{aguevara@perimeterinstitute.ca}  \affiliation{Perimeter Institute for Theoretical Physics, Waterloo, ON N2L 2Y5, Canada}\affiliation{Department of Physics and Astronomy, University of Waterloo, Waterloo, ON N2L 3G1, Canada}\affiliation{CECs Valdivia and Departamento de F\'isica, Universidad de Concepci\'on, Casilla 160-C, Concepci\'on, Chile} 
%
%
 \noaffiliation
\vskip 0.25cm